\documentclass[aps,preprint,amssymb,amsmath,eqsecnum,nofootinbib]{revtex4}
 \newtheorem{question}{QUESTION}[section]
 \begin{document}
 \title{Einstein's lifts and topologies: topological investigations on the Principle of Equivalence}
 \author{Gavriel Segre}
 \homepage{http://www.gavrielsegre.com}
 \email{info@gavrielsegre.com}
 \thanks{I would like to thank Tullio Regge for having shown me how the issue I was inquiring
 was linked with the subtilities of the boundary-conditions in the Cauchy-problem for Einstein's equation.
 I would then like to thank Vittorio de Alfaro  and Marco Cavagli\'{a} for many teachings on Constraints' Theory and Paolo Aschieri and Giovanni Landi
 for precious teachings on Noncommutative Geometry. They all have of course no responsibility as to any error contained in these pages.}
 \begin{abstract}
   The gedanken-experiment of Einstein's lift is analyzed in order
   of determining whether the free-falling observer inside the
   lift can detect the eventual topological non-triviality of
   space-time, as it would seem considering a non-globally-hamiltonian action of the symmetry
   group of the observer's action (that, unfortunately, can be obtained only
   submitting the lift also to a suitable electromagnetic field)
   and considering that the observer can locally detect the
   topological alteration of the constants-of-motion's algebra.

   It follows that a problem exists in formalizing  the Principle of
   Equivalence, owing to its  indetermination as to the topology of the reference's flat space-time
   defining the special relativistic laws to which, up to first
   order terms in the normal coordinates of the lift's Lorentz
   moving inertial frame, all the non-gravitational Laws of Physics
   have to collapse.

   It is then shown how the problem may be avoided getting rid of the Principle of
   Equivalence following the Hawking-Ellis' axiomatization of
   General Relativity purely based on the assumption of the
   Einstein-Hilbert's action.

   Connes' axiomatization of General Relativity having as
   only dynamical variable the spectrum of Dirac's operator is
   then used to discuss the initial topological question concerning
   Einstein's lift in the language of Spectral Geometry,
   explicitly showing its inter-relation with the celebrated
   Marc Kac's issue whether one can hear the shape of a drum, and
   showing how Index Theory is the natural framework in which some
   partial answer may be obtained.

   The whole issue is then analyzed in Connes' Quantum Gravity,
   suggesting how Noncommutative Geometry allows, through Noncommutative Index Theory,
   to get some insight along the footsteps followed in the
   commutative case.

   Some attempt of relating the issue to Anandan's claim on the
   difference among the holonomies of General Relativity and  the holonomies of
   Yang-Mills' theories is finally reported.
  \end{abstract}

 \maketitle
 \newpage
 \tableofcontents
 \newpage
\section{Local signature of the global topology} \label{sec:Local signature of the global topology}
Let us consider a particle of mass m free-falling in the
space-time $ ( M \, , \, g) $: it is described by a classical
dynamical system, that I will denote as  $FREE-FALL_{m} (M \, , \,
g) $, determined by the action functional:
\begin{equation}\label{eq:action functional of free-fall}
  S[ q^{\mu} ( \lambda ) ] \; := \; - \, m
  \int_{\lambda_{1}}^{\lambda_{2}} d \lambda \, \sqrt{ - g_{\mu \, \nu} ( q( \lambda ) ) \, \frac{ d q^{\mu} ( \lambda) }{ d \lambda } \frac{ d q^{\nu} ( \lambda) }{ d \lambda } }
\end{equation}
The reparametrization invariance of the paths in eq.\ref{eq:action
functional of free-fall} generates the existence of a primary
first class constraint \cite{Dirac-01},
\cite{Henneaux-Teitelboim-92} consisting in the
on-mass-shell-condition, expressed, adopting Penrose's abstract
index convention \cite{Wald-84}, as:
\begin{equation}\label{eq:on mass shell condition}
  H \; = \; p_{a} g^{a \, b} p_{b} \, + \, m^{2} \; \approx \; 0
\end{equation}
individuating the coisotropic submanifold
\cite{McDuff-Salamon-98}:
\begin{equation}
  C \; := \; \{ \, ( p \, ,  \, q) \: : \in \: T^{\star} M \: : \:
   p_{a} g^{a \, b} p_{b} \, + \, m^{2} \, = \, 0 \}
\end{equation}
of the phase space $ \Gamma \, := \, ( T^{\star} M \, , \,
\omega_{st} ) $ of $ FREE-FALL_{m} (M \, , \, g) $, where $
\omega_{st} $ is the standard symplectic form of M's cotangent
bundle.

Denoted by $ \sigma $ the 2-form induced by $ \omega_{st} $ on
the constraint's submanifold C:
\begin{equation} \label{eq:induced 2-form}
  \sigma \; := \; i^{^{\star}} \omega_{st}
\end{equation}
($ i \, : \, C \, \mapsto \, M $ being the inclusion),
Weinstein's reduction of the pre-symplectic manifold $ ( C \, , \,
\sigma ) $ individuates the reduced phase space:
\begin{equation} \label{eq:reduced phase space}
  \Gamma_{RED} \; := \; ( \frac{C}{K} \, , \, \omega )
\end{equation}
where $ K \, := \, T C^{\perp} $ is C's characteristic
distribution and $ \omega $ is the induced symplectic form on $
\frac{C}{K} $.

Let us now consider the time-like geodesic $ \gamma $ solution of
the Cauchy's problem:
\begin{equation}\label{eq:motion equation for free-fall}
  \frac{ \delta S [ q ( \tau ) ] }{ \delta q ( \tau )  } \;  = \; 0
\end{equation}
\begin{equation}\label{eq:initial condition for free-fall}
  q(0) \;  = \; q_{1}
\end{equation}
\begin{equation}\label{eq:final condition for free-fall}
  q(0) \;  = \; q_{2}
\end{equation}
where I have imposed the parametrization through proper-time
(proper-time gauge fixing) and where $ q_{1} $ and $ q_{2} $
belong to a geodesically convex open U of M.

Referring to the celebrated \emph{gedanken experiment} of
Einstein's lift, let us suppose our particle to be an observer
closed inside the walls of a lift, itself free-falling.

Let us now analyze M's topology:

if we assume Penrose's strong-form of the Cosmic Censorship
Conjecture stating the global hyperbolicity of $ ( M \, , \, g )
$, i.e. stating the existence in $ ( M \, , \, g ) $ of
Cauchy-surfaces, the topology of M is of the form $ {\mathbb{R}}
\, \times \, \Sigma $, $ \Sigma $ being any Cauchy 3-surface of $
( M \, , \, g ) $.

The problem of classifying the homeomorhism's classes of
topological-3-manifolds is (as I know) still open; this is
partially owed to the fact that Poincar\'{e}'s conjecture (that
every 3-homotopic-sphere $ S_{h}^{(3)} $ is homeomorphic to $
S^{(3)} $) is still open \footnote{I consider it open until the
Clay Mathematical Institute certifies that the recent proof by
M.J. Dunwoody \cite{Dunwoody-02} is the correct answer to
http://www.claymath.org/prizeproblems/poincare.htm}.

Let us observe that, fortunately,  the fact that a Cauchy
3-surface $ \Sigma $ is not simply a tridimensional topological
manifold but is endowed with a differentiable structure doesn't
change the game since every topological 3-manifold admits one and
only one differentiable structure \cite{Nash-94}.

Contrary, if we don't assume Penrose's strong-form of the Cosmic
Censorship Conjecture we have no restrictions on the topology of
M.

The problem of classifying 4-topological manifolds was proved by
Markov to be recursively-unsolvable still in 1958, the
undecidability deriving, essentially, from the
recursive-unsolvability of group-theoretic problems regarding the
fundamental groups of the 4-topological manifolds (cfr. the $
7^{th} $ chapter "Decision Problems" of
\cite{Collins-Zieschang-98})

The problem of classifying the simply-connected
4-topological-manifolds has been "almost solved" in 1982 by M.H.
Freedman \cite{Freedman-82} through a theorem stating that:
\begin{itemize}
  \item the simply-connected 4-topological-manifolds with even
  intersection form \cite{Nash-94}, \cite{Jost-95} are all
  homeomorphic
  \item  the simply-connected 4-topological-manifolds with odd
  intersection form \cite{Nash-94}, \cite{Jost-95} are divided in
  two equivalence-classes of homeomorhisms
\end{itemize}
Let us observe, furthermore, that in a space-time $ ( M \, , \, g
) $ M is a differentiable manifold; this complicates the game
since, in four dimensions, the topological-manifolds and the
differentiable-manifolds are no more bijective:
\begin{itemize}
  \item Donaldson \cite{Donaldson-83} has proved that, given a
  simply-connected 4-differentiable-manifold with
  positive-definite intersection form, such an intersection form
  is diagonalizable on the integers; an immediate corollary of
  Donaldson's Theorem is that a simply-connected
  4-topological-manifold having even and positive-definite
  intersection form doesn't admit differentiable structures.

 \item it is possible, anyway, to determine suitable hypothesis'
 under which a 4-topological manifold admits differentiable
 structures. E.g. Quinn \cite{Quinn-82} has proved that it
 is sufficient to require that M is not compact.
\end{itemize}
If a topological manifold of dimension greater than three admits
differentiable structures it doesn't imply, anyway, that it admits
only one differentiable structure. The dimension four is, as to
the classification of differentiable structures, a case for its
own: the existence, in certain cases, of a finite, discrete number
of exotic structures for compact topological manifolds of
dimension different from four (e.g. the 27 exotic-spheres
discovered by Milnor \cite{Milnor-56} in 1956) is a result of
Obstruction's Theory, i.e. it is corresponding to the
non-triviality of some characteristic class; the (eventual)
existence of exotic structure in four dimensions has different
origins and may also lead to an uncountable infinity of this kind
of structures.

E.g., as it has been proved by Gomft \cite{Gomft-85}, $
{\mathbb{R}} ^{4} $ admits an uncountable infinity of
differentiable structures.

The problem of classifying the topologies and the differentiable
structures of 4-topological-manifolds is, indeed, one of the most
fascinating research fields of contemporary Mathematical Physics,
owing to:
\begin{itemize}
  \item its link \cite{Freed-Uhlenbeck-84} with the structure of
  the Moduli-space $ Mod_{k} $ of the instantons of classical
  SU(2)-Yang-Mills-theories built on principal bundles with second
  Chern's class k (already displayed by the proof of Donaldson's
  Theorem)
  \item the introduction by Donaldson \cite{Donaldson-87},
  \cite{Donaldson-90} of a class of powerful topological
  invariants for simply connected 4-topological-manifolds, namely
  the celebrated Donaldson's polynomials (certain integer
  symmetric polynomials in the second integer homology)
  \item Witten's discovery \cite{Witten-88}, \cite{Witten-94} that
  Donaldson's polynomials may be obtained as correlation functions
  of a suitable BRST-supersymmetric topological field theory
  \item the introduction by Seiberg nd Witten \cite{Witten-94} of
  a simpler approach based on coupled equations for a section of
  a linear bundle and  a connection on an auxiliary fibre bundle
  with abelian gauge group (precisely U(1)).
\end{itemize}
In these note, anyway, I won't enter into the mathematical
sophistications concerning the classification of the topologies
of 4-differentiable- manifolds.

\bigskip

I will instead concentrate all the attention on two basic
physical questions:
\begin{enumerate}
  \item
\begin{question} \label{qu:local observability of the global topology}
\end{question}
   \textbf{QUESTION ON THE LOCAL OBSERVABILITY OF THE GLOBAL
  TOPOLOGY}

  \emph{can the observer in the lift understand, during the proper-time interval $ [ 0 \, , \, T ] $, if the topology of M is trivial ?}
  \item
\begin{question}  \label{qu:correct formulation of the Principle of Equivalence}
\end{question}
   \textbf{QUESTION ON THE CORRECT FORMULATION OF THE PRINCIPLE OF
  EQUIVALENCE}

  \emph{supposing that the question \ref{qu:local observability of the global topology}
  has affirmative answer, can this fact be formalized saying that
  Mechanics of the dynamical system $ FREE-FALL_{m} ( M \, , \, g
  ) $ is obtained applying the Principle of Equivalence by taking
  as reference space-time not the topologically trivial minkowskian space-time $  (
  {\mathbb{R}}^{4} \, , \, \eta := \, - d x^{0} \bigotimes d x^{0}
   \, + \, d x^{i} \bigotimes d x^{i} ) $, but a topologically
   non-trivial modification of it, i.e. a space-time of the form $  (
  X \, , \, \eta := \, - d x^{0} \bigotimes d x^{0}
   \, + \, d x^{i} \bigotimes d x^{i} ) $ where X is a
   topologically non-trivial 4-differentiable-manifold ?}
\end{enumerate}
\newpage
\section{Non-globally-hamiltonian actions of the isometries on the reduced phase
space} \label{sec:Non-globally-hamiltonian actions of the
isometries on the reduced phase space} In this section I will try
to discuss question\ref{qu:local observability of the global
topology} using a celebrated result of Analitical Mechanics: the
cohomological interpretation of the existence of actions
 by symplectic diffeomorphisms of symmetry
groups on the phase space of a classical dynamical system such
that they are not globally-hamiltonian, i.e. they haven't an
equivariant momentum map \footnote{unfortunately not all the
authors follow the terminology by Marsden and Ratiu that I have
adopted \cite{Marden-Ratiu-94}: e.g. Marsden and Ratiu's notion
of globally-hamiltonianity is called simply hamiltonianity by Mc
Duff and Salamon \cite{McDuff-Salamon-98} while it is called
poissonianity by Arnold and Givental \cite{Arnold-Givental-01}}.

Given the classical dynamical system $ FREE-FALL_{m} (M \, , \, g
) $ let us consider an action-from-left $ \Phi $  through
symplectic diffeomorphisms of the isometries's group G of $ ( M
\, , \, g ) $ on the reduced phase-space $ \Gamma_{RED} \; :=  (
\frac{C}{K} \, , \, \omega ) $, i.e.:
\begin{equation}
  \Phi \, : \, G \times \Gamma_{RED} \, \mapsto \, \Gamma_{RED}
  \; :  \;
  ( \Phi_{g} \in Diff(\Gamma_{RED}) \, and \, ( \Phi_{g} )^{\star}
  \omega \, = \, 0 ) \; \; \forall g \in G
\end{equation}
Obviously G is a symmetry group of $ FREE-FALL_{m} (M \, , \, g )
$, i.e.  its motion-equation is invariant under $ \Phi $.

A moment map of $ \Phi $:
\begin{equation}
   \vec{J}  \, : \,   \Gamma_{RED} \, \mapsto \, L^{\star} (G) \: : \: <
   \vec{J} (x) \, , \, A > \: = \: J_{A} (x)
\end{equation}
(where $ J_{A}  \in C^{\infty} ( \Gamma_{RED} ) $  is the
classical observable generating the one-parameter group of
symplectic diffeomorphisms  $ \{ \Phi_{\exp ( t \, A)} \} $) is
equivariant, i.e. such that:
\begin{equation}
  \{  J_{A} \, , \, J_{B} \} \; = \; J_{[ A \, , \, B ]} \; \;
  \forall \, A , B \, \in  \, L(G)
\end{equation}
if and only if the 2-cocycle $ \Sigma \in Z_{2} [ L(G) ] $:
\begin{equation}
  \Sigma ( A \, , \, B ) \; := \; T_{e} \sigma_{B} (A) \; \;
  \forall \, A \, , \, B \in L(G)
\end{equation}
is identically null, where the map $ \sigma_{A} \, : \, G \,
\mapsto \, {\mathbb{R}} $:
\begin{equation}
  \sigma_{A} (g) \; := \; \sigma (g) \; \; \forall \, A \, \in \,
  L(G)
\end{equation}
is defined through the map $ \sigma \, : \, G \, \mapsto \, Hom (
L(G) \, , \, {\mathbb{R}}) $:
\begin{equation}
  \sigma(g) \, \cdot \, A \; := \; P_{A , g}
\end{equation}
defined, in its turn, through the family of maps $ \{  P_{A , g}
\, : \, \Gamma_{RED} \, \mapsto \, {\mathbb{R}} \, , \, A \in L(G)
\, , \, g \in G \} $:
\begin{equation}
  P_{A,g} (x) \; := \; < \vec{J} ( \Phi_{g} (x)) \, , \, A > \, -
  \, < Ad^{\star}_{g^{- 1}} \, , \, A >
\end{equation}
where $ Ad^{\star} $ is the co-adjoint representation of G.

The cohomology class $ [ \Sigma ] \in H^{2} [L(G)] $ is
univocally determined by the action $ \Phi $.

Let us suppose that $ \Sigma $ is not null.

If $ [ \Sigma ]  \; = \; [ 0 ] $ it is possible, anyway, "to
repair" (through a suitable re-definition of the $ \{ J_{A} \, ,
\, A \in L(G) \} $ corresponding to the addition to $ \Sigma $ of
a 2-coborder) the momentum map $ \vec{J} $ in order to obtain a
different momentum map having the equivariance property; the
action is, conseguentially, globally hamiltonian.

If $ [ \Sigma ]  \; \neq \; [ 0 ] $, contrary, no "repairing" of
$ \vec{J} $ allows to result in a new equivariant momentum map.
Such a situation may be described as a classical anomaly: an
equivariant momentum map may be obtained only at the price of
substituting G with a central extension of its in complete
analogy with the situation occurring in quantum anomalies
\cite{De-Azcarrega-Izquierdo-95}, where the impossibility of
representing  a symmetry group in a non-projective way, again
owed to the non-triviality of a suitable 2-cocyle, may be
"repaired" only constructing an ordinary representation of a
central extension.

When $ H^{2} [L(G)] \; = \; 0 $, as it  happens, for example, if
$ ( M \, , \, g ) \; = \;   (
  {\mathbb{R}}^{4} \, , \, \eta := \, - d x^{0} \bigotimes d x^{0}
   \, + \, d x^{i} \bigotimes d x^{i} ) $, every action $ \Phi $
   of G on the reduced phase space is globally hamiltonian.

We can then ask ourselves the following:

\begin{question} \label{qu:local observability of a nonequivariant map}
\end{question}

\textbf{QUESTION ON THE LOCAL OBSERVABILITY OF THE NON-EQUIVARIANT
MAP}

\emph{can the observer in the lift notice the alteration of the
Lie algebra of the motion's constants owed to a non-equivariant
momentum map ?}

\bigskip

If the answer to question\ref{qu:local observability of a
nonequivariant map} is affirmative, it is then evident the
importance, for the resolution of the question\ref{qu:local
observability of the global topology}, of the following:

\begin{question} \label{qu:non-triviality of Killing algebra implies
non-triviality of space-time}
\end{question}
\textbf{QUESTION ON THE POSSIBILITY OF INFERRING THE TOPOLOGICAL
NON-TRIVIALITY OF A SPACE-TIME FROM THE
COHOMOLOGICAL-NON-TRIVIALITY OF THE ALGEBRA OF ITS KILLING VECTOR
FIELDS}

\emph{is it possible, from the observation of the topological
non-triviality of $ H^{2} [L(G)] $ to infer the topological
non-triviality of M ?}

\bigskip

Indeed if the answers to both question\ref{qu:local observability
of a nonequivariant map} and question\ref{qu:non-triviality of
Killing algebra implies non-triviality of space-time} were
positive, this would imply that the answer to
question\ref{qu:local observability of the global topology} would
be positive too.
\newpage
\section{Correct formalization of the Principle of Equivalence} \label{sec:Correct formalization of the Principle of Equivalence}
In this paragraph I will  investigate which implication an
eventual positive answer to question\ref{qu:local observability
of a nonequivariant map} and question\ref{qu:non-triviality of
Killing algebra implies non-triviality of space-time} would have
as to question\ref{qu:local observability of the global topology},
as to question\ref{qu:correct formulation of the Principle of
Equivalence} and, ultimativelly, on the issue concerning the
correct formalization of the Principle of Equivalence.

I will, consequentially, assume that the observer closed in the
lift may, through a measurement of the constants of motion and of
the algebraic relations they satisfy, notify a suitable
topological non-triviality of M during the short proper-time's
interval $ [ 0 \, , \, T ] $.

Let us consider a Cartan's gauge \cite{Spivak-79}, i.e. a local
section of the general frame bundle $ GLM ( M \, , \, GL(4,
{\mathbb{R}} ) )$ and let us assume that:
\begin{itemize}
  \item it is adapted to the timelike geodesic arc $ \gamma $, defined by the Cauchy problem of
eq.\ref{eq:motion equation for free-fall}, eq.\ref{eq:initial
condition for free-fall} and eq.\ref{eq:final condition for
free-fall}
  \item all the frames in the set:
\begin{equation}
  s_{\gamma} \; := \; \{ ( e_{0} (x) \, , \, \cdots \, , e_{3} (x)
  ) \: : \: x \in \gamma \bigcap M^{s} \} \; \subset \; s
\end{equation}
(where $M^{s}$ denotes the definition's domain of s) are
orthonormal
  \item all the frames' elements $ e_{0} (x) \, , \, x \in \gamma
  $ are equal to the tangent vectors of the geodesic for a
  suitable choice, let us call it $ \lambda $, of the affine
  parameter
\end{itemize}
Such a Cartan's gauge is sometimes called  \cite{Prugovecki-92},
\cite{Prugovecki-95} a Lorentz-moving-inertial-frame for $ \gamma
$.

Let us now construct the normal coordinates associated to s; this
may be done by the following steps:
\begin{enumerate}
  \item let us pose $ X^{0} ( \lambda ) \, = \, \lambda $ in all
  the points along $ \gamma $
  \item  for every point $ x \in \gamma $ let us draw all
  space-like geodesics with tangent vectors:
\begin{equation}
  X \; = \; X^{i} e_{i} (x) \;  \in \; T_{x} M
\end{equation}
in  a fixed parametrization
  \item in a suitable neighborhood of $ x( \lambda ) \in \gamma $
  let us assign to every point y belonging to one of these
  geodesics at unitary parametric distance from $ x( \lambda) $
  the coordinates $ ( X^{0} \, , \, X^{1}, \, , \, X^{2} \, , \,
  X^{3} ) $
  \item making this operation for every point of $ \gamma $ we
  obtain a coordinates' system $ ( X^{0} \, , \, X^{1}, \, , \, X^{2} \, , \,
  X^{3} ) $, defined in a suitable tube around $ \gamma $, that we
  define to constitute the normal coordinates associated to the
  Lorentz moving inertial frame s for $ \gamma $.
\end{enumerate}
We can now formalize  the Principle of Equivalence in the
following way \footnote{ Though not agreeing on Prugovecki's
ideas on Quantum Gravity, I adopt here his formulation of the
Principle of Equivalence strongly demanding to
\cite{Prugovecki-92}, \cite{Prugovecki-95} for all the underlying
conceptual sophistications such as as Friedman's distinction
between first-order-laws and second-order-laws \cite{Friedman-83}
and its rule as to the Factor-Ordering Problem (cfr. the section
16.3 "The Factor-Ordering Problems in the Principle of
Equivalence" of \cite{Misner-Theorne-Wheeler-73})} :

\bigskip

\textbf{PRINCIPLE OF EQUIVALENCE:}

\emph{for every Lorentz moving inertial frame s free-falling
along a time-like geodesic $ \gamma $ all the non-gravitational
laws of Physics, expressed in the normal coordinates associated to
s, have in each point of $ \gamma $ to be equal, up to
first-order in these coordinates, to the corresponding
special-relativistic laws expressed in the coordinates associated
to the respective Lorentz frames}

\bigskip

Let us now consider our observer who, in the assumed hypotheses,
is able to detect, during the short proper-time interval $ [ 0 \,
, \, T ] $ and without looking out of the lift, the topological
non-triviality of M.

Using the Principle of Equivalence in the way opposite to the one
usually adopted, i.e. utilizing the knowledge of a
non-gravitational law on $ ( M \, , \, g ) $ to infer the form of
the corresponding special-relativistic law, we may infer that the
algebraic relations defining the constants-of-motion's Lie
algebra in the special-relativistic dynamical system
corresponding to $ FREE-FALL_{m} ( M \, , \, g ) $ must
themselves show the same alteration that, for $ FREE-FALL_{m} ( M
\, , \, g ) $, shows the topological non-triviality of M.

This fact may be rephrased saying that the special relativistic
dynamical system corresponding to $ FREE-FALL_{m} ( M \, , \, g )
$ is not the dynamical system $ FREE-FALL_{m} (
  {\mathbb{R}}^{4} \, , \, \eta := \, - d x^{0} \bigotimes d x^{0}
   \, + \, d x^{i} \bigotimes d x^{i} ) $ but a
   special-relativistic dynamical system of the form $ FREE-FALL_{m} (
  X \, , \, \eta := \, - d x^{0} \bigotimes d x^{0}
   \, + \, d x^{i} \bigotimes d x^{i} ) $, where X is some
   topologically non-trivial 4-differentiable-manifold.

   Under the assumed hypotheses this could be at its turn rephrased
   saying that \textbf{the reference-spacetime to use to generalize the
   non-gravitational special-relativistic laws to the
   topologically non-trivial space-time $ ( M \, , \, g ) $ is
   not the Minkowskian space time $ (
  {\mathbb{R}}^{4} \, , \, \eta := \, - d x^{0} \bigotimes d x^{0}
   \, + \, d x^{i} \bigotimes d x^{i} ) $ but a flat space-time $ (
  X \, , \, \eta := \, - d x^{0} \bigotimes d x^{0}
   \, + \, d x^{i} \bigotimes d x^{i} ) $ where X is some
   topologically non-trivial 4-differentiable-manifold}.

   \bigskip

We have, up to this point, assumed that it may be the case that
the momentum map for the action of the isometries' group on the
reduced phase is not equivariant; unfortunately this is not the
case, requiring a slight technical modification of the previous
analysis.

The key point consists in the Theorem of Global Hamiltonianity
for Cotangent Lifts \cite{Marden-Ratiu-94} whose formulation
requires the introduction of some preliminary notion.

Given two differentiable-manifold $ Q_{1} $ and  $ Q_{2} $ and a
diffeomorphism $ f \, : \, Q_{1} \mapsto Q_{2} $, the cotangent
lift of f is the map $ T^{\star} f \, : \, T^{\star} Q_{1} \,
\mapsto \, T^{\star} Q_{2} $ defined as:
\begin{equation}\label{eq:cotangent lift}
  < \, T^{\star} f ( \alpha_{s} ) \, , \, v \, > \; := \; < \,
  \alpha_{s} \, , \, f_{\star} v \, > \; \; \alpha_{s} \in
  T_{s}^{\star} Q_{1} \, , \, v \in T_{q} Q_{2} \, , \, s = f(q)
\end{equation}
Let us now consider a differentiable-manifold Q on which a
left-action $ \Phi $ of a Lie group G by diffeomorphisms is
defined.

This action may then be associated with an a action $
\Phi_{\star} $ of G on the symplectic manifold $ (T^{\star} Q \,
, \, \omega_{st}) $, called the cotangent-lift (left) action of $
\Phi $, defined as:
\begin{equation}
  ( \Phi_{\star} )_{g} \; := \; T^{\star}_{g \, \cdot \, q} ( \Phi_{g^{-
  1}})
\end{equation}
The Theorem of Global Hamiltonianity for Cotangent Lifts states
that any cotangent-lift (left)-action is globally hamiltonian.

It should be clear what a calamity this theorem is for our
purposes: it implies that the action of the isometries'-group of
a space-time $ ( M \, , \, g ) $ on the reduced phase-space $
\Gamma_{RED} $ has an equivariant momentum map, so that it cannot
be useful  as to our search of an example allowing to give a
positive answer to question\ref{qu:local observability of a
nonequivariant map}.

Fortunately there exist a way of bypassing this obstacle,
consisting in considering not a free-particle, but a particle
minimally-coupled with an instanton $ \nabla $, belonging to the
moduli space with topological charge $ Mod_{k} $, of a Yang-Mills
theory on $ ( M \, , \, g ) $ with gauge group G (assumed to be a
compact and connected Lie group)
\cite{Balachandran-Marmo-Skagerstam-Stern-83}, \cite{Nash-94},
\cite{Jost-95}.

If $ k \; = \; 0 $ the principal bundle $ P( M \, , \, G ) $ of
the Yang-Mills theory is trivial, i.e. $ P \; = \; M \, \times \,
G $, and  hence admits a global fibre chart. Conseguentially $
\nabla $ admits a global potential A. In this case we can easily
infer that, again, the reparametrization invariance of our
particle's action leads to the existence of a primary first class
constraint stating the weak-vanishing of the hamiltonian:
\begin{equation}
  H \; := \; ( p_{\mu} \, - \, A_{\mu} ) g^{\mu \nu} ( p_{\nu} \, - \, A_{\nu}
  ) \; \approx \; 0
\end{equation}
If, contrary, $ k \; \neq \; 0 $, $ P(M \, , \, G) $ is
non-trivial and, conseguentially, it doesn't admit a global fibre
chart so, that the instanton $ \nabla $ cannot be described by a
single global potential.

Let us then take in account a family of fibre-charts $ \{ ( U_{i}
\, , \, \varphi_{i} ) \} $ such that $ \{ U_{i} \} $ is a
contractible open covering of M, and  the associated family of
local one-potentials $  \{ A_{i} \} $. For every chart we may
consider the local hamiltonian:
\begin{equation}
  H_{i} \; := \; ( p_{\mu} \, - \, A^{(i)}_{\mu} ) g^{\mu \nu} ( p_{\nu} \, - \, A^{(i)}_{\nu}
  )
\end{equation}
undefined outside $ U_{i} $.

These considerations would then lead to infer that, for $ k \;
\neq \; 0 $, it is not possible to obtain a single global
hamiltonian. This is indeed true until we insist in requiring
that the phase-space's symplectic structure is the standard one.

Altering the symplectic form in order to absorb the interaction
with the Yang-Mills field into the phase-space's  geometric
structure, it is, anyway, possible to define the particle's
classical dynamical system in a way completely independent on the
potentials of $ \nabla $.

Demanding to the literature \cite{Guillemin-Sternberg-84} for
more detailed informations, it will be enough here to mention that
it is possible to define a functional $  T-Mod_{k} \; : \; Mod_{k}
\; \mapsto \; \Omega^{2} ( T^{\star} M ) $, that I will call a
modular-term, with the following properties:
\begin{enumerate}
  \item $ \omega_{st} \; + \; T-Mod_{k} $ is a symplectic form
  over $ T^{\star} M $
  \item the dynamical system of our particle minimally-coupled with
  the instanton $ \nabla $ may be defined as the classical
  dynamical system with phase-space $ ( T^{\star} M \; , \;
  \omega_{st} \, + \, T-Mod_{k} ) $ and hamiltonian $ H \; = \;
  p_{a} g^{a b } p_{b} \, + \, m^{2} $.
\end{enumerate}
In the electromagnetic case $ G \, =  \, U(1) $ a particular
modular term is given by:
\begin{equation}
  T-Mod_{k} \; := \; \pi^{\star} F_{\nabla}
\end{equation}
where $  F_{\nabla} \; := \; \nabla \, \circ \, \nabla $ is the
curvature of the instanton $ \nabla $.

Let us now observe that the alteration of the phase-space's
symplectic structure "neutralizes" the  Theorem of Global
Hamiltonianity for Cotangent Lifts.

This implies  that, by adding the minimal interaction with the
fixed electromagnetic  field, it is possible to generate
non-equivariant momentum maps.
   \newpage
   \section{The indetermination of the Principle of Equivalence and the possibility of avoiding it in axiomatizing  General
   Relativity} \label{sec:The indetermination of the Principle of Equivalence and the possibility of avoiding it in axiomatizing  General Relativity}
The trial of furnishing an axiomatization of General Relativity
has engaged a lot of people for eigthy years.

General Relativity  was founded by Einstein on two principles
\cite{Einstein-Lorentz-Weyl-Minkowski-52}:
\begin{enumerate}
  \item the Principle of Equivalence
  \item the Principle of General Covariance
\end{enumerate}
Yet in 1917, anyway,  E. Kretschmann \cite{Kretchmann-17}
objected  that the Principle of General Covariance is
tautological, because   any putative physical law may be written
in a way that make it hold good for all system of coordinates.
Following this observation, most of the following axiomatizations
descarded the Principle of General Covariance and founded the
whole theory on the Principle of Equivalence alone; this is, more
or less explicitely , the attitude of almost all the more popular
manuals on General Relativity, such as that by Weinberg
\cite{Weinberg-70}, that by Misner, Thorne and Wheeler
\cite{Misner-Theorne-Wheeler-73}, and that by  Wald
\cite{Wald-84} \footnote{Weinberg introduces the Principle of
General Covariance in the section 4.1 of \cite{Weinberg-70} just
to simplify the analysis, but states that it is a consequence of
the Principle of Equivalence and reports Kretschmann's observation
on its tautological nature. Misner, Thorne and Wheeler discuss its
rule in the section 17.6 of \cite{Misner-Theorne-Wheeler-73}
asserting that Mathematics was not sufficientely refined in 1917
to distinguish among the demand for "no prior geometry" and the
demand for a "geometric, coordinate-independent formulation of
physics, encoding both in the condition of "General Covariance
whose  ambiguous nature is seen as the source of all the
confusions concerned with its discussions from Kretschmann and
beyond. Wald introduces it in the section 4.1 of \cite{Wald-84}
stating it as the condition that there are no preferred vector
fields or preferred bases of vector fields pertaining only to the
structure of space which appear in any law of physics. He then
remarks its vagueness owed to the fact that the phrase
"pertaining to space" does not have a precise meaning.}.

As I have shown in the last section, anyway, the correct
formalization of the Principle of Equivalence is highly
not-determined, not specifying the topology of the reference-flat
lorentzian space-time  $ (
  X \, , \, \eta := \, - d x^{0} \bigotimes d x^{0}
   \, + \, d x^{i} \bigotimes d x^{i} ) $ involved as far as \emph{"the corresponding special relativistic
   laws"} are concerned.

Fortunately, it is possible to get rid of such an ambiguity
corncerning  the Principle of Equivalence, adopting a different
axiomatization of General Relativity in which the Principle of
Equivalence takes no parts: the axiomatization by Hawking and
Ellis \cite{Hawking-Ellis-73} based on:
\begin{enumerate}
  \item the Principle of Local Causality
  \item the Principle of  Local Energy Conservation
  \item the Principle of the Einstein-Hilbert Action
\end{enumerate}
While the Principle of Local Causality and the Principle of Local
Energy Conservation pose some constraint on the  energy-momentum
tensor of matter-fields and hence on the action $ S_{matter} [
\phi \, , \, g_{a b }] $ describing them, the Principle of the
Einstein-Hilbert Action states that the action  describing the
gravitational field is the functional $ S_{gravity} \; : \;
Lor_{4}(X) \, \mapsto \, {\mathbb{R}} $:
\begin{equation}\label{eq:Einstein-Hilbert action}
  S_{gravity} [ g_{ab} ] \; := \;  N _{EH} \, \int d \mu [g_{a \,  b} ] \, R[
  g_{a b}]
\end{equation}
where $ N _{EH} $ is a  $ g_{ab}$-independent normalization
factor, X is a 4-differentiable-manifold and $Lor_{4}(X) $ is the
set of all 4-lorentzian metrics on it.

General Covariance, formalized in a suitable way, may then be seen
as a theorem deriving from the Hawking-Ellis' axioms, as we will
discuss more extensively in section\ref{sec:On the locality of
General Relativity and of Gauge Theories}

The problems  concerning the topological indetermination of the
reference space-time X in the Principle of Equivalence becomes, in
this framework, completely enclosed in the issue concerning the
boundary conditions in the Cauchy Problem for the minimum-action's
principle for the whole action:
\begin{equation}
  S[ g_{a  b} \, , \, \phi ] \; := \; S_{gravity} [ g_{ab} ] \, +
  \, S_{matter} [ \phi \, , \, g_{ab}  ]
\end{equation}
\newpage
\section{Can the observer inside Einstein's lift  hear the topology of space-time
?} \label{sec:Can the observer inside Einstein's lift  hear the
topology of space-time ?}

In a celebrated article in 1966  \cite{Kac-66} Marc Kac formulated
the following classical problem of Spectral Geometry that I report
in the Protter's formulation cited by Gilkey \cite{Gilkey-95} (a
detailed exposition of Kac's article is available in the fifth
chapter "Spectral Geometry with operators of Laplace type" of
\cite{Esposito-98}):
\begin{center}
  \emph{"Suppose a drum is being placed in one room and a person with perfect pitch hears but cannot see the drum.
  Is it possible for her to deduce the precise shape of the drum just from hearing the fundamental tone and all the overtones
  ?} (cited by the section 4.2 "Isospectral manifolds" of
  \cite{Gilkey-95})
\end{center}
I will now show  how Kac's question is similar to our
question\ref{qu:local observability of the global topology}:

Let us observe, first of all, that the observer closed inside
Einstein's Lift has access only to local information exactly as
the observer of Kac's problem.

And he tries to use such local information to infer the global
geometrical structure of the space(-time) in which he lives.

But as we will see the analogy goes further, since the same kind
of local information is of the same type: the spectrum of a
suitable operator.

 At this purpose let us observe that, under the mild assumption
that the first two Stiefel-Whitney classes  $ w_{1} (TX) \;  , \;
w_{2} (TX) $ of the 4-differentiable-manifold X vanish, it is
possible to give an alternative axiomatization of General
Relativity in which the Principle of the Einstein-Hilbert Action
is replaced by  the Principle of the Connes Action
\cite{Connes-94}, \cite{Connes-98}, \cite{Landi-97},
\cite{Landi-Rovelli-98}, \cite{Gracia-Bondia-Varilly-Figueroa-01}.
stating that the action  describing the gravitational field is
the functional $ S_{gravity} \; : \; SPIN-SPECTRA \, \mapsto \,
{\mathbb{R}} $:
\begin{equation} \label{eq:Connes action}
  S_{gravity}  [  Sp (D_{g}) \, , \, \Lambda ] \; := \; N_{C} \, Tr ( \chi ( \frac{D^{2}
  }{\Lambda^{2}}))
\end{equation}
($ N _{C} $ being a $Sp (D_{g})$-independent normalization
factor), where:
\begin{equation}
   SPIN-SPECTRA \; := \; \{ Sp (D_{g}) \, , \, g \in
   Riem(X) \}
\end{equation}
where $ D_{g} $ is the Dirac operator of the spin-manifold $ ( X
\, , \, g ) $, $ Riem(X) $ is the set of all the riemannian
metrics over X, $ \Lambda $ is a cut-off and $ \chi $ is a
suitable cut-off function throwing away the contribution of all
the eigenvalues of $ D_{g} $ greater than $ \Lambda $.

It is important to remark that the key point  giving foundation
to the substitution of the Principle of the Einstein-Hilbert
Action with the Principle of the Connes' Action as to the
axiomatization of General Relativity is the theorem stating that
the category having as objects the closed finite-dimensional
spin-manifolds and as morphisms their diffeomorphisms is
equivalent to the category having as objects the abelian spectral
triples and as morphisms their automorphisms.

It is useful, at this point, to introduce the following
terminology:

given a riemannian manifold $ (X \, , \, g ) $ let us define its
spectrum  as the spectrum of the Laplace-Beltrami operator on it;
furthermore, given a spin-manifold $ (X \, , g ) $,  let us
define its spin-spectrum as the spectrum of the Dirac operator on
it.

We can now precisely state Kac's question in the following way :
\begin{question} \label{qu:determination of a riemannian manifold by its spectrum}
\end{question}
QUESTION IF A RIEMANNIAN MANIFOLD IS DETERMINED BY ITS SPECTRUM
\begin{center}
  \emph{is a riemannian manifold  $ (X \, , g ) $ determined by its spectrum ?}
\end{center}
We can then state the same question as to the Dirac operator:
\begin{question} \label{qu:determination of a spin-manifold by its spectrum}
\end{question}
QUESTION IF A SPIN-MANIFOLD IS DETERMINED BY ITS SPIN-SPECTRUM
\begin{center}
  \emph{is a spin-manifold  $ (X \, , g ) $ determined by its spin-spectrum ?}
\end{center}
That one cannot hear the shape of a drum, i.e. that the answer to
both question\ref{qu:determination of a riemannian manifold by its
spectrum} and question\ref{qu:determination of a spin-manifold by
its spectrum} is negative, was proved by the determination of
 different isospectral riemannian-manifolds and spin-manifolds.

It must be observed, anyway, that our question\ref{qu:local
observability of the global topology} is much less ambitious,
asking only if:
\begin{question} \label{qu:determination of the topology of a spin-manifold by its spectrum}
\end{question}
QUESTION IF THE TOPOLOGY OF A SPIN-MANIFOLD IS DETERMINED BY ITS
SPIN-SPECTRUM
\begin{center}
  \emph{is the topology of a spin-manifold  $ (X \, , g ) $ determined by its spin-spectrum ?}
\end{center}
The natural framework in which to discuss
question\ref{qu:determination of the topology of a spin-manifold
by its spectrum} is Index Theory:

by the Atiyah-Singer Index Theorem the index of the Dirac
operator $ D_{g} $ over the spin-manifold $ ( X \, , \, g ) $:
\begin{equation}
  Index( \, D_{g} ) \; := \; dim( Ker(D_{g}) ) \, - \,  dim(Coker(D_{g}))
\end{equation}
may be expressed as:
\begin{equation}\label{eq:Atiyah-Singer-index-theorem for the Dirac operator}
  Index(D_{g}) \; = \;  \int_{M} \hat{A} (X)
\end{equation}
where:
\begin{equation} \label{eq:A-genus}
  \hat{A} (X) \: = \; \prod_{i=1}^{4} \frac{\frac{x_{i}}{2}}{\sinh(\frac{x_{i}}{2})}
\end{equation}
is the $ \hat{A} $-genus of X expressed in terms of the
eigenvalues $( \, x_{1} \, , \, x_{2} \, , \, x_{3} \, , \, x_{4}
\, ) $ of the block-form diagonalization of the curvature 2-form
\cite{Alvarez-95}:
\begin{equation}
  {\mathcal{R}}_{a b} \; = \; \begin{pmatrix}
    0 & x_{1} & 0 & 0 & 0 & 0 & 0 & 0 \\
    - x_{1} & 0 & 0 & 0 & 0 & 0 & 0 & 0 \\
    0 & 0 & 0 & x_{2} & 0 & 0 & 0 & 0 \\
    0 & 0 & - x_{2} & 0 & 0 & 0 & 0 & 0 \\
    0 & 0 & 0 & 0 & 0 & x_{3} & 0 & 0 \\
    0 & 0 & 0 & 0 & - x_{3} & 0 & 0 & 0 \\
    0 & 0 & 0 & 0 & 0 & 0 & 0 & x_{4} \\
    0 & 0 & 0 & 0 & 0 & 0 & - x_{4} & 0 \
  \end{pmatrix}
\end{equation}

Since the  $ \hat{A} $-genus of  X is a characteristic class,
eq.\ref{eq:Atiyah-Singer-index-theorem for the Dirac operator}
implies that the index of the Dirac operator is a topological
invariant.

Given two spin-manifolds $ ( \, X_{1} \, , \, g_{1} \, ) $ and $
( \, X_{2} \, , \, g_{2} \, ) $ the fact that:
\begin{equation}
   X_{1} \, \sim \, X_{2} \; \Rightarrow \; Index(D_{g_{1}}) \, =
  \, Index(D_{g_{2}})
\end{equation}
doesn't unfortunately imply the converse:
\begin{equation}
  Index(D_{g_{1}}) \, = \, Index(D_{g_{2}}) \; \nRightarrow \; X_{1} \, \sim \,
X_{2}
\end{equation}
doesn't allowing to infer completely the space-time's topology
from the dimensions of  Dirac-operator's kernel and cokernel.

Though not determining it, anyway,
eq.\ref{eq:Atiyah-Singer-index-theorem for the Dirac operator}
shows how Index Theory allows to extract information concerning
the space-time's topology from the spectrum of Dirac's operator,
that, as we have seen, may be considered the only dynamical
variable of General Relativity.
\newpage
\section{Noncommutative Einstein's lifts and  noncommutative
topologies} \label{sec:Noncommutative Einstein's lifts and
noncommutative topologies}

In this section I will briefly sketch how all the analysis' made
in the previous sections may be generalized to the quantum case.

Since, according to Noncommutative Geometry \cite{Connes-94},
\cite{Connes-98}, \cite{Landi-97}, \cite{Landi-Rovelli-98},
\cite{Gracia-Bondia-Varilly-Figueroa-01}, a quantum spacetime is
nothing but a noncommutative spectral triple $ ( A \, , \,
{\mathcal{H}} \, , \, D )$  we can introduce the noncommutative
analogue of Einstein's-lift's gedanken-experiment (that I will
denote briefly as the noncommutative-lift's gedanken-experiment)
 and state the noncommutative analogue of question\ref{qu:local
observability of the global topology}:
\begin{question} \label{qu:local observability of the global noncommutative topology}
\end{question}
   \textbf{QUESTION ON THE LOCAL OBSERVABILITY OF THE GLOBAL
  NONCOMMUTATIVE TOPOLOGY}

  \emph{can the observer in the noncommutative lift understand, during the proper-time interval $ [ 0 \, , \, T ] $, if the noncommutative topology of $ ( A \, , \,
{\mathcal{H}} \, , \, D )$  is trivial ?}

trying to discuss it in terms of the noncommutative analogue of
question\ref{qu:determination of the topology of a spin-manifold
by its spectrum}, i.e. of the following:
\begin{question} \label{qu:determination of the topology of a noncommutative  by its spectrum}
\end{question}
\textbf{QUESTION IF THE TOPOLOGY OF A NONCOMMUTATIVE SPACE-TIME IS
DETERMINED BY ITS SPIN-SPECTRUM}
\begin{center}
  \emph{is the topology of a noncommutative space-time $ ( A \,
  , \, {\mathcal{H}} \, , \, D )$ determined by its spin-spectrum ?}
\end{center}
whose precise meaning lies on the two basic theorems of
Noncommutative Topology:
\begin{enumerate}
  \item the Gelfand-Naimark's Theorem stating the equivalence among
  the category of  the compact Hausdorff topological spaces and the category of the abelian $ C^{\star}$-algebras
  \item the Serre-Swan's Theorem stating the equivalence among
  the category of vector bundles over a compact topological space
  M and the category of the finitely generated projective modules
  over C(M)
\end{enumerate}
The analysis of the previous section strongly suggest that the
natural framework in which one can try to get some insight into
question\ref{qu:determination of the topology of a
noncommutative  by its spectrum} is Noncommutative Index Theory.

Now, despite some non-rigorous, naive folklore concerning
noncommutative-coordinates at Planck's scale
\cite{Doplicher-Fredenhagen-Roberts-95}, \cite{Doplicher-01} and
some even more muddling folklore \cite{Longo-01} discussing
Quantum Index Theory in the wrong  framework of Algebraic Quantum
Field Theory, Noncommutative Index Theory is a well defined
subject of Noncommutative Geometry  with well-established theorems
such as the Noncommutative Atiyah-J\"{a}nich's Theorem or the
Connes-Moscovici's Local Index Formula
\cite{Gracia-Bondia-Varilly-Figueroa-01}.
\newpage
\section{On the locality of General Relativity and of Gauge
Theories} \label{sec:On the locality of General Relativity and of
Gauge Theories}

On discussing the issue concerning the compatibility among the
locality of General Relativity and the claimed non-locality of
Quantum Mechanics \footnote{I don't agree both on the often
claimed fact (assumed by Anandan) that the Aharonov-Bohm effect
proves the non-locality of Quantum Mechanics \cite{Healey-99}
(the Aharonov-Bohm effect being  simply a particular application
of Chern-Simons' topological quantum field theory \cite{Hu-01})
and on the even more often claimed statement \cite{Shimony-00}
that the quantum mechanical violation of Bell's inequalities is a
proof of the (claimed) quantum non-locality: the quantum
violation of Bell's inequalities is nothing but an immediate
consequence of the fact that Noncommutative Measure Theory, the
ground floor of Noncommutative Geometry, is irreducible to
Commutative Measure Theory \cite{Connes-94}, \cite{Streater-00a}}
Jeeva Anandan has recently  \cite{Anandan-02} returned to develop
a claim he made years before \cite{Anandan-93}: the fact that
that the locality of General Relativity lies on the fact that its
similarity with the (claimed)  non-local Yang- Mills' theories is
only partial.

The discussion of Anandan's claim may be useful to clarify better
how the  the Hawking-Ellis' axiomatization of General Relativity

\begin{itemize}
  \item allows, through the Principle of Local Causality, to give up the
topological indetermination I have shown to infect the Principle
of Equivalence
  \item is on  no way in contradiction with the possibility of locally
  hearing some (partial) information about the topology of space-time, revealing how the
  "global versus local issues"  are subtle when both K-theory and the correct formulation of Principle of Local Causality are taken
  into account
\end{itemize}
The starting point of Anandan's analysis is Trautman's discussion
of the following:
\begin{question} \label{qu:if General Relativity Theory is a Yang-Mills theory}
\end{question}
\textbf{QUESTION IF GENERAL RELATIVITY IS A GAUGE THEORY}

\emph{Is General Relativity a Yang-Mills' theory of a suitable
principal bundle P(M, G), for a suitable choice of the base space
M and the structure group G ?}

\bigskip

The first attempts to give a positive answer to question\ref{qu:if
General Relativity Theory is a Yang-Mills theory} were made by
Einstein and Dirac through their investigations in the forthies on
the possibility of formulating General Relativity in a way such
that the dynamical variable is a connection, as  in Yang-Mills'
theories, and not a metric \cite{Gambini-Pullin-96}.

The natural starting point appeared, with this respect, to be the
substitution of the Einstein-Hilbert's action of
eq.\ref{eq:Einstein-Hilbert action} with the  Palatini's action:
\begin{equation}\label{eq:Palatini's action}
  S_{gravity} [ g_{a b} \, , \, \nabla_{a} ] \; := \; \;  N _{EH} \, \int d \mu [g_{a \,  b} ] \, R_{a b}[
  \nabla_{a}] g^{a b}
\end{equation}
in which the connection $ \nabla_{a} $ is taken as an independent
dynamical variable, fixed to be the Levi-Civita connection of $
g_{a b } $ by the dynamical equation obtained, together with
Einstein's equation, varying  eq.\ref{eq:Palatini's action} w.r.t.
to both $ \nabla_{a} $ and $ g_{a b} $.

This would lead to suspect that General Relativity is nothing but
the Yang Mills theory w.r.t. a principal bundle $ P( X , SO(1,3)
) $, where X is some 4-differential manifold.

I will know show, anyway, that such a suspicion is wrong.

The group of gauge transformations for an $ SO( 1 \, , \, 3 )
$-Yang Mills theory on X is defined as the group of vertical
automorphisms of the underlying principal bundle $ P( X , SO(1,3)
) $ given by::
\begin{equation}
  {\mathcal{G}}_{YM} \; := \; \Gamma ( X \, , \, Ad \, P )
\end{equation}
where:
\begin{equation}
  Ad \, P \; := \; P \, \times_{Ad SO(1,3)} \, SO(1,3)
\end{equation}
is the associated adjoint bundle of $ P( X , SO(1,3) ) $.

The group of gauge transformation for General Relativity is,
contrary, the group of automorphims of GLM preserving its
\textbf{soldering form} $ \theta $ \cite{Trautman-80},
\cite{Prugovecki-92}:
\begin{equation} \label{eq:theorem of general covariance}
  {\mathcal{G}}_{GR} \; = \; \{ \alpha \in  Aut (GLM) \; : \, \alpha
  \circ \theta  \, = \,  \theta \}
\end{equation}
containing no other vertical automorphism than the identity:
\begin{equation} \label{eq:difference in the group of gauge trasnformations}
  {\mathcal{G}}_{GR} \, \bigcap \, {\mathcal{G}}_{YM} \; = \; {\mathbb{I}}
\end{equation}
Since:
\begin{equation}
   {\mathcal{G}}_{GR} \; = \; Diff(M)
\end{equation}
eq.\ref{eq:theorem of general covariance} is nothing but the
Theorem of General Covariance, namely the correct mathematical
formalization of the Principle of General Covariance, avoiding
the tautological bug of its naive formulation shown by Kretchmann
that we discussed in section\ref{sec:The indetermination of the
Principle of Equivalence and the possibility of avoiding it in
axiomatizing General Relativity}.

It is importance to remark again  that General Covariance is a
theorem and not a principle of General Relativity, being implied
by the Hawking-Ellis' axiomatization we adopted:

eq.\ref{eq:theorem of general covariance} is implied by the
Diff(M)-invariance of  eq.\ref{eq:Einstein-Hilbert action} and,
hence, by the Principle of the Einstein-Hilbert's Action.

As a consequence a space-time is represented mathematically not
by a 4 lorentzian manifold $ ( M \, , \, g_{a b} ) $ but by an
element of the quotient space $ \frac{Lor(M)}{Diff(M)} $,
automatically getting rid of all the confusions concerning the
\emph{"hole-argument"} of the 1913 Einstein-Grossmann's paper
\cite{Stachel-89}, \cite{Norton-89}, \cite{Prugovecki-92},
\cite{Prugovecki-95}.

\smallskip

The negative answer to question\ref{qu:if General Relativity
Theory is a Yang-Mills theory} resulting from Trautman's analysis
\footnote{It must be remarked, with this respect, that, defining
a a gauge theory as a generic physical theory having as dynamical
variable a connection on a principal bundle, Trautman's
conclusion is that, though not being a Yang-Mills' theory,
General Relativity is a gauge theory;  we are using here, anyway,
a more restrictive definition of a  gauge theory as a synonimous
of a Yang-Mills' theory}  reflects, according to Anandan, the
following radical structural difference:
\begin{itemize}
  \item Gauge theories are nonlocal, since its holonomies are nonlocal
  objects, invariant under change of the enclosed "flux" by one
  "quantum"
  \item General Relativity is local since, owing to the peculiar
  rule of the \textbf{soldering form} having no analogous in gauge theories,
  its holonomies are local objects, noninvariant under change of the enclosed "flux" by one
  "quantum"
\end{itemize}
Though intuitivelly suspecting that Anandan's remark could be
stricly connected with the issue discussed in the sections
\ref{sec:Local signature of the global topology},
\ref{sec:Non-globally-hamiltonian actions of the isometries on
the reduced phase space},  \ref{sec:The indetermination of the
Principle of Equivalence and the possibility of avoiding it in
axiomatizing  General Relativity}, \ref{sec:Can the observer
inside Einstein's lift  hear the topology of space-time ?} I have
not succeeded yet in formalizing such a link.

A first step in this direction would consist in analyzing
Anandan's issue in the Ashtekar's formulation of General
Relativity \cite{Ashtekar-91}, \cite{Gambini-Pullin-96} whose
phase space and canonical variables are exactly those of a
(complex) SU(2)-Yang-Mills theory, while its reduced-space is a
subspace of the reduced phase-space of a (complex) SU(2)
Yang-Mills theory (obtained taking the quotient w.r.t. the Gauss
Law) owing to the existence of four further constraints.

As to the quantum case  discussed in
section\ref{sec:Noncommutative Einstein's lifts and
noncommutative topologies} the natural framework to analyze its
interrelation with Anandan's remark would consist in comparing
the loop representations of a quantum complex SU(2) theory and
Loop Quantum Gravity \cite{Rovelli-91b}, \cite{Gambini-Pullin-96}
based on  the loop representation of Ashtekar's formalization of
General Relativity as to the rule of the soldering form and the
locality of holonomies \footnote{The fact that Loop Quantum
Gravity differs from Connes' Quantum Gravity and that both these
theories differ from String Quantum Gravity shouldn't, in my
modest opinion, be dramatized: different approaches to Quantum
Gravity should be seen as attempts of climbing a mountain from
different faces}.

\end{document}